\documentclass[10pt]{IEEEtran}

\makeatletter
\def\ps@headings{%
\def\@oddhead{\mbox{}\scriptsize\rightmark \hfil \thepage}%
\def\@evenhead{\scriptsize\thepage \hfil \leftmark\mbox{}}%
\def\@oddfoot{}%
\def\@evenfoot{}}

\makeatother

\usepackage{amssymb,amsmath}
\usepackage{amsfonts}
\usepackage{array}
\usepackage{algorithmicx}
\usepackage[ruled,commentsnumbered]{algorithm2e}
\usepackage{subfigure}
\usepackage{mathrsfs}
\usepackage[T1]{fontenc}
\usepackage[utf8]{inputenc}
\usepackage{textcomp}
\usepackage{eurosym}
\usepackage{graphicx}
\usepackage{multirow}
\usepackage{epstopdf}
\usepackage{epsfig}
\usepackage{stmaryrd}
\usepackage{multirow,url}
\usepackage{color,cite}
\usepackage{slashbox}
\usepackage{amsthm}
\usepackage{bm}

\usepackage{multirow}
\usepackage{caption}

\begin{document}

% paper title
\title{
SDN-enabled Tactical Ad Hoc Networks: Extending Programmable Control to the Edge}

 \vspace{-5mm}
\author{\IEEEauthorblockN{Konstantinos Poularakis$^\ddagger$, George Iosifidis$^{*}$, Leandros Tassiulas$^\ddagger$}
\\
\IEEEauthorblockA{$^\ddagger${Department of Electrical Engineering and Institute for Network Science, Yale University, USA}\\
{$^{*}${School of Computer Science and Statistics, Trinity College Dublin, Ireland}}
}
 \vspace{-5mm}
}

\maketitle
\begin{abstract}
Modern tactical operations have complex communication and computing requirements, often involving different coalition teams, that cannot be supported by today's mobile ad hoc networks. To this end, the emerging Software Defined Networking (SDN) paradigm has the potential to enable the redesign and successful deployment of these systems. In this paper, we propose a set of novel architecture designs for SDN-enabled mobile ad hoc networks in the tactical field. We discuss in detail the challenges raised by the ad hoc and coalition network environment, and we present specific solutions to address them. The proposed approaches build on evidence from experimental evaluation of such architectures and leverage recent theoretical results from SDN deployments in large backbone networks.
\end{abstract}

\thispagestyle{empty}

\section{Introduction}\label{sec:intro}

Today's tactical communications involve convoluted and dynamic patterns of information circulation or processing, and raise challenges that are not encountered in commercial systems. Indeed, we currently lack the necessary tools for designing communication systems that can efficiently support the full spectrum of tactical missions. Moreover, the distance between state-of-the-art solutions and the military needs grows fast, as the latter become increasingly complex. This fact has driven Department of Defense (DoD) agencies to repeatedly and emphatically outline the priorities of tactical communications, calling both academia and industry for actions towards a research breakthrough.

\subsection{The Operational Requirements}
One of the main requirements for modern military communication systems is to support tactical field operations in \emph{areas without infrastructure}. A method that has been long considered as the ultimate solution to achieve this goal is the deployment of Mobile Ad hoc Networks (MANETs). These are envisioned as fully decentralized systems with self-organizing capabilities, hence having the required robustness and scalability. Nevertheless, the MANETs that are currently deployed at the edge, function at a basic level, and suffer from issues such as complex configuration requirements and protocol overhead due to network topology changes.

An additional requirement that arose recently for tactical MANETs is the increasing need to support \emph{coalition operations}. This latter term describes communication between teams of soldiers (or, other actors in the tactical field) that belong to different command centers or even different nations. The role of a coalition network is to enable the circulation of information, such as mission data, while ensuring that sharing of sensitive information is no wider than necessary, which is known as the ``need-to-know'' (NTK) principle. Despite the army efforts in this area, e.g., the Multilateral Interoperability Program run by several nations, we lack mechanisms that achieve the necessary coalition interoperability.

\begin{figure*}[t]
		\vspace{-2mm}
	\begin{center}
		\includegraphics[scale=0.6]{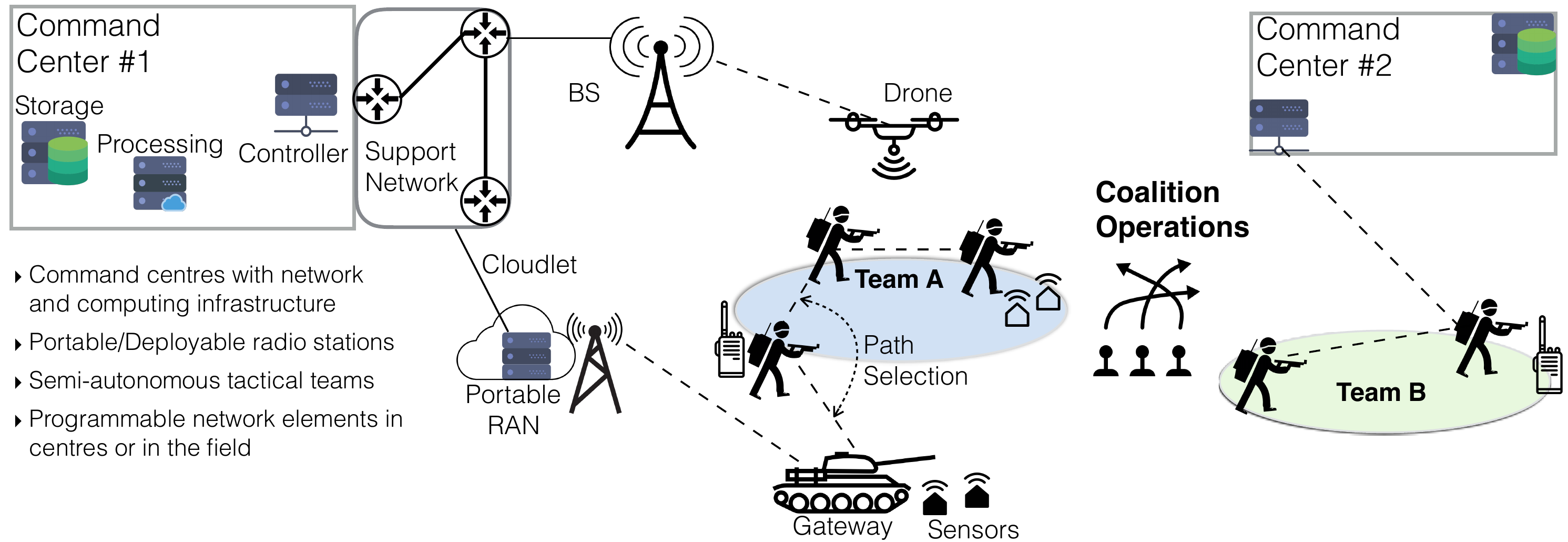}
		\small{\caption{\textbf{Modern Tactical Environment}. Tactical communication networks include edge-networking components, multi-hop and multi-path data flows, and coalition networks to support cross-team communications.}}
%		Some nodes can be enhanced with SDN capabilities (denoted by "S"), in-network processing can be supported in cloudlets and other fast deployable infrastructures, and this in turn can facilitate SDN deployment at the network edge.
		\label{fig:overall}
	\end{center}
	\vspace{-2mm}
\end{figure*}

%\vspace{-1mm}
\subsection{The Opportunity}
The above operational requirements at the tactical level raise technical issues which are very challenging to address. At the same time, the networking community witnesses today the advent of Software Defined Networking. SDN constitutes a paradigm shift as it enables a full or partial decoupling of network control and data forwarding operations, and brings unprecedented programmability to network management \cite{sdn-commag}.
Therefore, a question that unavoidably arises is: \emph{How can we leverage SDN to design tactical ad hoc networks}?

At a first glance, SDN seems a promising solution for enabling the deployment of tactical MANETs. First, the currently prevalent approach for completely decentralized MANET architectures is, likely, one of the main reasons they are not used at large. This philosophy can be revisited with SDN, which offers centralized control and network-wide view, and indeed there are already suggestions for employing SDN in MANETs \cite{syrivelis}. Moreover, in coalition tactical operations SDN can facilitate flexible routing and dynamic transport-level decisions, thus enabling information circulation based on mission-specific criteria and the NTK constraints. This opportunity has been identified by military and industry experts \cite{dinesh}. Nevertheless, currently it is not clear which architectures are suitable for these SDN-enabled mobile networks, what are the key system parameters that affect their performance, and how we can optimize them. 

In this paper we make a first step towards addressing, in a systematic fashion, the above issues. First, we address the design of SDN-enabled MANETs, or \emph{SMANETs}, as the next generation of tactical ad hoc networks.
We provide the blueprint of a SMANET, including the system architecture for the mobile nodes. The foremost issue here is to decide \emph{where to place and how to organize the SDN control logic (controllers) in the network}. These decisions essentially shape the performance of the SMANET.
%For example, placing controllers closer to mobile nodes increases the MANET's agility but might overload the nodes that host the controllers.
%On the other hand, syncing frequently the forwarding elements with distant-located controllers might consume significant bandwidth and introduce delays.
We explain how these decisions can be optimized, and we provide supporting evidence from testbed evaluations of such ad hoc schemes.

% (controller placement and responsibilities are fixed)
Next, we focus on the data plane, i.e., the data forwarding nodes. Here, the following two issues are important. First, tactical networks, and especially coalition networks, will often comprise a large number of heterogeneous network elements. For example, some teams or soldiers therein, may not have SDN-enhanced radio nodes. This will result in hybrid systems where SDN data plane nodes co-exist with legacy nodes that use non-SDN routing protocols. The key technical question here is to decide \emph{where to deploy the SDN forwarding elements and how to use them}. Second, current SDN proposals rely on the centralized controller to update all forwarding rules at the data plane nodes. In tactical networks with high level of dynamism and frequent network failures, this centralization will result in slow network updates, as well as significant controller overhead. Hence, a second question is \emph{how to make data plane nodes to autonomously react to network changes, but at the same time preserve the benefits of centralized control}. We explain that these decisions are heavily affected by the operational needs, involve intricate optimization problems and we present approaches for tackling them.

The rest of this paper is organized as follows: in Section \ref{sec:background} we provide some background on tactical and SDN research.
%we discuss the operational requirements in modern tactical missions and technical communication challenges.
In Section \ref{sec:controllers} we propose a novel architecture for SDN-enabled MANETs, analyze their performance limits, and discuss various control plane design approaches. Section \ref{sec:coalitions} focuses on the data plane of the tactical network, considering scenarios of hybrid SDN deployment and semi-autonomous data forwarding elements, and provides methodologies for optimizing its operation. We conclude and discuss the road ahead in Section \ref{sec:conclusions}. 
\section{Background and Current State}\label{sec:background}

In this section, we discuss the operational requirements in modern tactical missions and the technical communication challenges. 

\subsection{Modern Tactical Operations}

Modern military missions include a large number of diverse actors (soldiers, vehicles, etc.) who may operate independently, in coordination with each other, or in sync with command centers. The latter are fixed or deployable infrastructures, often airborne (e.g., drones), that work in a semi-autonomous fashion serving the actors or bridging them with higher-level commands. The actors might be organized in different groups, e.g., belonging to different commands, that have independent or intertwined goals. An instance of a modern tactical environment and the envisaged communication support is depicted in Figure 1.

%Each actor may have multiple network interfaces that must be seamlessly managed so as to select the more suitable communication path for each task.

\begin{figure*}[t]
	\begin{center}
		\includegraphics[scale=0.7]{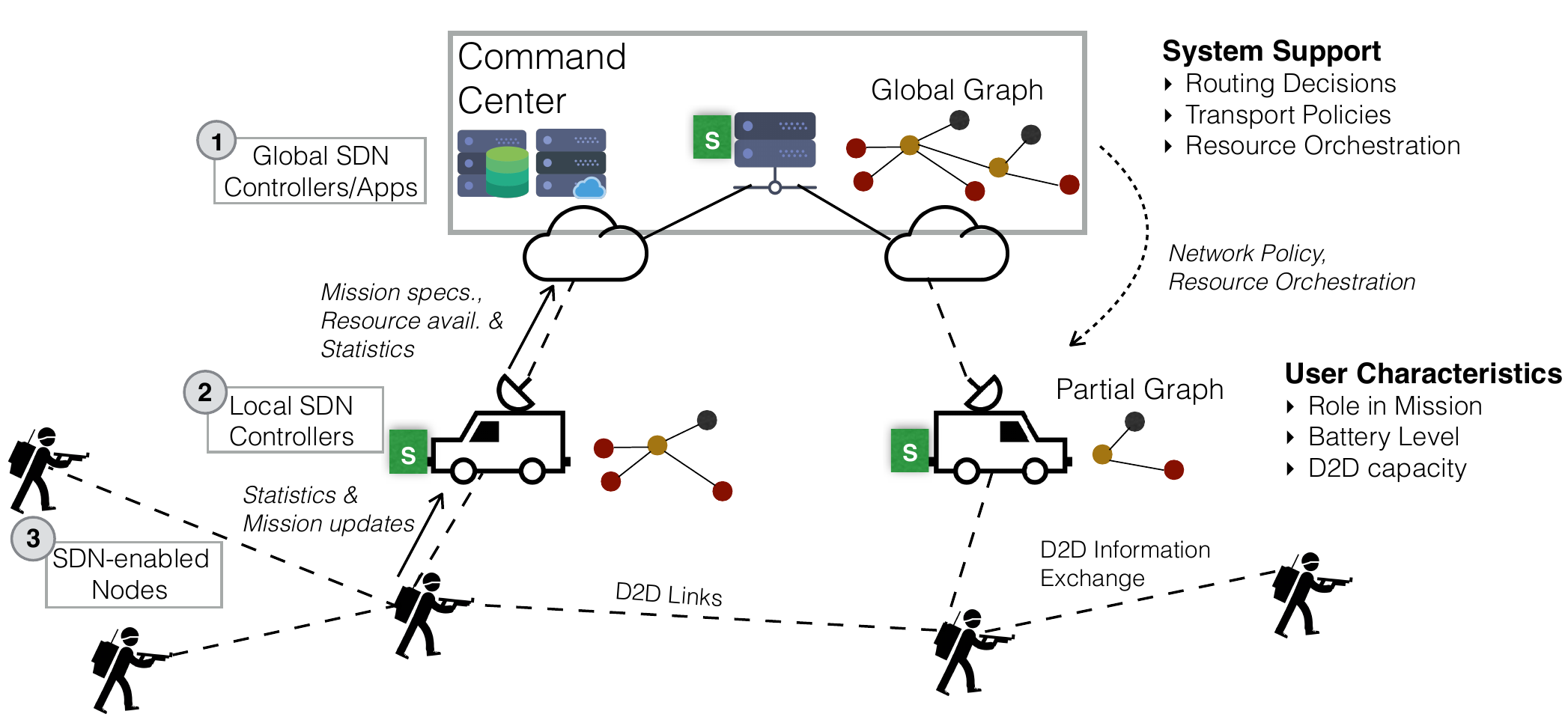}
		\small{\caption{\textbf{Blueprint of the proposed SDN-enabled MANET architecture}. SDN local controllers (denoted by "S") receive the network policies from the cloudlet (global controller) and configure accordingly the end-nodes under their supervision.}}
		\label{fig:blueprint}
	\end{center}
\end{figure*}

Military missions increasingly rely on timely information which has to be collected through sensors, and processed in the tactical field, e.g., in handheld devices or in the cloud.  Besides, DoD agencies such as DARPA have recently called for the development of mission-aware computation and communication systems\footnote{See, for example, \emph{Dispersed Computing} DARPA-BAA-16-41, and \emph{Content-Based Mobile Edge Networking}, DARPA-BAA-11-51, as well the Federated Mission Networking (FMN) concept by NATO.}. They also emphasize the importance of programmable networks that enable dynamic in-network decisions, such as path selection and bandwidth provisioning. Due to these requirements, MANETs are even more important today as they need to support field ad hoc communications and also provide multi-hop connectivity to command centers.

Moreover, the equipment in modern tactical missions includes heterogeneous network and computing elements, such as various mobile radios and field sensors. These systems have particular constraints, e.g., operate on tight energy budgets, and this introduces various performance versus lifetime trade-offs. At the same time, this equipment often operates in adversarial environments and hence needs to be robust in link and other failures, as well as to several types of security attacks.

\subsection{MANETs and Coalition Operations: What is missing}

One of the main challenges in MANETs is efficient routing, and various protocols have been proposed to address it, e.g., the Optimized Link State Routing (OLSR). These solutions can result in intermittent connectivity, long-lasting disconnections or may generate excessive protocol overhead due to frequent topology changes. Moreover, they are not flexible enough to adapt to the new operational requirements described above. Delay-tolerant networking has been proposed to address the connectivity problem, but this covers only a narrow slice of the military missions. The common denominator of these prior efforts is the focus on fully decentralized architectures which is perceived as a prerequisite for achieving the necessary level of robustness. However, with the full or partial separation of control and data planes, SDN can provide the level of network view and centralized control that was missing to implement complex mission-critical applications.% Besides, MANETs today need to also provide connectivity to command centers and this new approach can serve this purpose as well.

Coalition networks is a relatively new operational requirement. The main problems that have been studied so far are mobility support, interoperability and scalability \cite{coalition-survey}. The traditional routing protocol used for the communication between coalition members is the Border Gateway Protocol. BGP was initially designed to exchange reachability information between autonomous systems (AS) in wireline networks. However, BGP cannot support mobility, nor has the adaptability that is necessary for the tactical domain. A way to prevent network fragmentation is by using a virtual router for each AS domain, but this raises scalability issues~\cite{coalition-virtual}. 
Besides, prior proposals for selecting the gateways and ensuring cross-domain secure connections, albeit very important, cannot cope with fast changing missions nor support detailed NTK policies. SDN brings the network programmability dimension that can potentially overcome these obstacles.

\subsection{The Advent of SDN}

There is a fast increasing volume of research in the area of SDN \cite{sdn-commag}. The vast majority of these studies refer to wireline infrastructures, i.e., ISP networks and data centers. A technical issue that has received considerable attention is the design of programmatic languages aiming to develop interfaces that are fully-flexible and scalable. This will enable network administrators to configure the data plane as needed, and design application-specific traffic control policies.

\begin{figure*}[t]
	\begin{center}
		\includegraphics[scale=0.8]{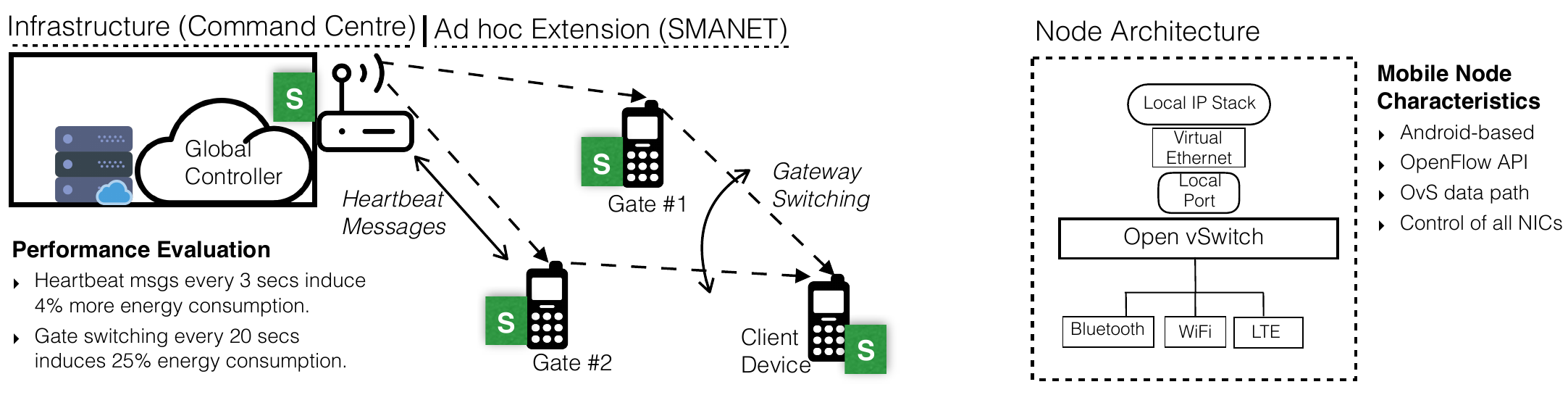}
		\small{\caption{\textbf{Performance and node architecture of SDN-enabled MANETs}. Left: The global controller assigns the routing rules to the local controllers that translate them to forwarding actions. Delays and energy consumption overheads are measured for bulky data delivery over multihop connections. Right: Node architecture, where an OpenFlow switch is installed at Linux-based nodes and takes under its control all network interfaces. Technical details can be found in \cite{syrivelis}.}}
		\label{fig:cones}
	\end{center}
\end{figure*}
%The main service runs the mission logic and communicates via the Northbound API with the global SDN controller

Few recent works have attempted to apply these ideas to wireless networks. For example, \cite{softcell} proposed to exploit SDN and shift control functions from core gateways to middle boxes. This can eliminate management bottlenecks by decentralizing the network operation. On the other hand, \cite{softran} suggested the deployment of software defined radio access networks (RAN). The idea is to assign the management of multiple base stations to a global controller and with this unified control improve performance. An interesting point is the suggestion for splitting the control decisions to those requiring full information (hence assigned to the global controller) and to those that need fast response (assigned to the radio elements). Going a step further to the data plane, \cite{openradio} proposed to turn the base stations to fully programmable nodes facilitating this way the virtualization of network resources.

The above developments manifest that small, yet solid steps have been made towards designing softwarized wireless networks. Interestingly, there have been also efforts to deploy SDN soft switches even to handheld devices, see \cite{syrivelis} and references therein. This manifests the actual potential for deploying SDN-enabled ad hoc networks. However, as it is explained below, there are many open issues that we need to address if we wish to apply these ideas to tactical ad hoc networks.

\section{SMANET Architecture Design}\label{sec:controllers}

This section discusses the design of SDN-enabled MANETs, or \emph{SMANETs}. We present below a set of key questions that arise when a SMANET architecture is designed, and provide suggestions for addressing them in a systematic way. This means that the architecture is tailored to the specific operational requirements, environment conditions, and capabilities of the equipment.

An example blueprint of the proposed SMANET architecture is illustrated in Figure 2. Here, two local SDN controllers are installed on portable stations that are in proximity with the mobile nodes. Each of them is able to view only a part of the network graph, and has to collect and send the respective network state information to the higher-level SDN controller located at a cloudlet. This global controller will construct the universal graph topology, devise the operation policies and disseminate them to the forwarding nodes through the local controllers. The latter can serve as bridges, or even take full control when the infrastructure connection is disrupted, or the data path queries need to be served in real time.

A key challenge in SMANETs is to determine the SDN controller architecture. This requires to decide both the \emph{placement} and \emph{organization} of controllers in the network. Regarding placement, a straightforward option is to place one controller to manage all the data forwarding devices in the network. Schemes that place many controllers have been proposed in the context of wired networks (e.g., see \cite{cp1} and \cite{cp2}). Regarding organization, the controllers need to be organized in such a way so as to be able to coordinate their actions in collecting network statistics, maintaining a consistent view of the network state (topology, traffic load, etc.) and sharing the burden of data path query service. For example, in a flat organization, all the controllers  maintain the same network state and can serve all the applications~\cite{hyperflow}. In a hierarchical controller organization, as in Fig. 2, however, different controllers will be able to serve different applications.

It is clear that in SMANETs the controllers can be placed in several locations such as the command centers, portable wireless infrastructure, or the mobile equipment.
%These decisions shape the performance of SMANETs and thus determine the missions they can support.
A proper approach to make these design decisions is by considering the pros and cons of each architecture.
Namely, having a central controller at a cloud server allows to configure all devices at once, but raises scalability concerns and may introduce non-negligible delays for policy updates. Therefore, for small groups of soldiers, and when there is strong connectivity with the controller, this is a proper solution.
On the other hand, placing the controller on a field platform allows SMANETs to control faster the forwarding elements (lower latency), to monitor the usage of wireless resources in real time, and mitigate interference. However, this might induce problems of consistency as the local controller might take locally-optimal but globally-inefficient decisions.
Thus, finding a controller placement that ensures the fair trade-off between the different objectives is needed~\cite{cp2}. A third option is to place the controllers at mobile nodes (e.g., the soldiers' equipment) so as to increase autonomicity and robustness to infrastructure link failures. However, this option should be carefully considered as the overheads might drain these resource-constrained devices.

\begin{figure*}
	\begin{center}
		\includegraphics[scale=0.6]{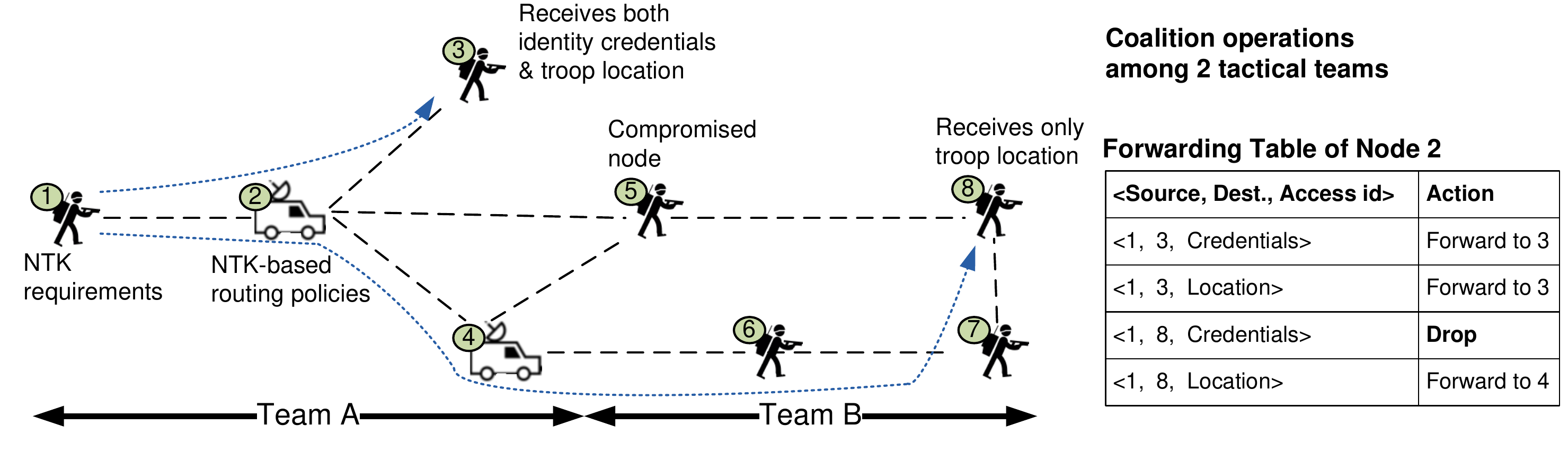}
		\small{\caption{\textbf{A coalition network example}. Two tactical teams form an SDN-enabled mobile ad hoc network. The example focuses on the forwarding table of node 2, which employs access \emph{ids} so as to manage information according to the NTK principle.}}
		\label{fig:coalition}
	\end{center}
\end{figure*}

At this point, it is important to quantify the above trade-offs. To this end, we can take lessons from recent results in mobile SDN systems. In particular, in our recent work \cite{syrivelis}, we experimented with SDN-enhanced portable Android devices coordinated by a global controller located in a clouldet. The goal was to provide a proof-of-concept prototype for a cloud-controlled ad hoc network where certain mobile devices will act as gateways, connecting other nearby client-nodes or aggregating multiple flows to satisfy throughput-hungry services. Extensive trials showed that such systems are technically feasible but have certain limitations and might induce important overheads.

Regarding the reconfigurability of the SDN-enhanced devices, we observed that an end-user (client device) can switch between gateways very fast, i.e., even every 20 seconds, while having an active connection with a distant server, e.g., for downloading a large file such as a map. The experimental set up is outlined in Figure 3. Nevertheless, more frequent reconfigurations consume significant battery energy (more than $20\%$, compared to the no-configurations benchmark) and induce non-negligible delays (more than $25\%$). On the other hand, updating the global controller (which was located within 1Km distance) with the current network status, did not consume significant device resources and could be even as frequent as every 3 seconds. Obviously, the actual numbers depend on a set of parameters, e.g., the number of hops for each flow, the physical distance of the devices, and so on. However, it is already clear which are the performance-cost trade-offs that should be carefully considered when one designs such architectures.

Clearly, a systematic optimization approach is needed to design the controller architecture in SMANETs.
%These decisions will be issued for long time periods, since controller migration requires significant overhead for state synchronization and device re-assignment.
A particular useful toolbox for devising placement policies is facility location theory. However, typical algorithms do not directly apply because of the practical constraints arising in SMANETs. Hence, the controller placement has to be casted as a discrete optimization problem that takes into consideration the topology of the network (available positions and their reach to forwarders), the applications and the network policies that realize them, the overhead of communicating the policies to data path, the cost for circulating the state synchronization messages among controllers (which depends on the controller organization), security and other special requirements of tactical operations, and the battery energy limitations of the devices. The controller placement can be also combined with other practical methods that increase network lifetime such as equipping a part of the nodes with power banks.

%Another important criterion that we should satisfy when designing the control logic of SMANETs is resilience. The SDN controller represents a single point of failure, which makes the architecture vulnerable to attacks or other outages. One way to avoid this issue is {replication}, where multiple controller instances manage the same part of the network rather than a single one (see \cite{disco}). Nevertheless, replication has been designed for data centers, and may not respond well to tactical environments. The overhead for communication between the replicas will be significant considering the scarcity of bandwidth in the tactical environment. Therefore, one should proactively decide the controllers' locations so as to facilitate their synchronization for replication purposes. 

\section{SMANET Data Plane Optimization}\label{sec:coalitions}

The data plane of a SMANET can benefit significantly by optimal design approaches. Consider for example the network in Figure 4 with eight SDN-enabled data plane nodes (soldiers and vehicles). The members of the two coalition teams should be able to exchange critical battlefield information, such as identity credentials and troop locations, but Team B members should have access only to the location data (because of NTK constraints). Using conventional protocols, like BGP and its extensions for tactical ad hoc networks \cite{coalition-virtual}, it is difficult to enforce such detailed routing policies. With SDN however, forwarding rules can match on a variety of header fields and not just the destination address. Hence, bespoken policies can be implemented to support the operational requirements. In the above scenario for example, one could install rules that forward packets based on the triplet (source id, destination id, access id), where the flag \emph{access id} captures the NTK requirements for each packet. This way, node 2 will forward identity credential packets only when they are destined to a Team A node; otherwise they will be dropped.

The above example highlighted the benefits of the extra design space brought by SDN in implementing detailed information exchange policies. 
An important, yet challenging, task is how to implement even more complex policies~\cite{softcell}. In the rest of this section, we emphasize two factors that make this task even more challenging in the context of tactical networks.

\begin{figure*}
	\begin{center}
		\includegraphics[scale=0.65]{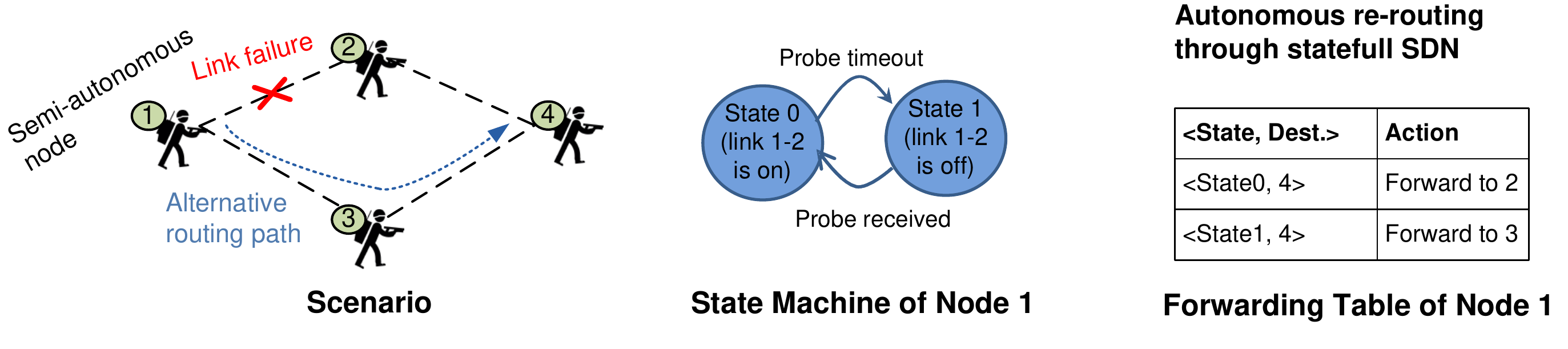}
		\small{\caption{\textbf{Statefull data plane example}. Node 1 autonomously re-routes traffic through the alternative path when detects that the link to node 2 has failed. This is realized by tracking the state of the link (using a state machine) and applying forwarding rules that match packets on current state.}}
		\label{fig:statefull}
	\end{center}
\end{figure*}

\subsection{Factor 1: Hybrid SDN Deployment}

In tactical networks, SDN data plane nodes will often co-exist (interfacing through Eastbound APIs) with legacy nodes that use standard legacy routing protocols. An important question in such hybrid systems is to determine where to deploy the SDN forwarding elements, e.g., which soldiers or vehicles should be equipped with SDN switches. These decisions are constrained by various factors. For example, each team may have only a certain number of SDN radios (budget), or during a mission there might exist limited time-windows for executing (or, reorganizing) such equipment upgrades.

Each SDN deployment decision impacts not only the node itself (the one being upgraded) but its 1-hop neighbors and even more distant nodes. To understand this, let us focus again on Figure 4. When node 2 is the only node that is upgraded to SDN, the NTK requirements for the flows emanating from node 1 can be satisfied. However, it is not possible with this deployment to satisfy additional requirements such as dynamically deciding not to route traffic through node 5 (e.g., when node 5 becomes suspicious of being compromised). Node 2 can override the OLSR shortest path (i.e., the path (2,5,8) with the minimum hop length) by routing the packets destined to node 8 through node 4 instead of node 5. In this case, the packets will follow the shortest path connecting node 4 with 8, i.e., the path (4,5,8). Nevertheless, both nodes 2 and 4 need to be upgraded to SDN in order to be able to route traffic through the secure path (2,4,6,7,8).

For the SDN deployment we can, to some extent, leverage results from previous studies that considered this problem in ISP backbone networks. In particular, in our recent work \cite{pist}, we presented greedy-based approximations which maximize the number of routing paths that are dynamically selectable though the SDN nodes. Our methodology was based on the theory of submodular functions. While this constitutes a first step in addressing the SDN deployment problem, SMANETs bring novel challenges. For example, in certain scenarios the actors not only move, but change drastically their relative positions. This might render previous SDN deployment decisions sub-optimal, and a re-arrangement might be necessary. %In such cases, reducing the cost of equipment redistribution is a vital objective.
%Similarly, the traffic control in tactical environments is more challenging as the information flow can be unpredictable in certain instances. Hence, previous solutions, e.g., \cite{lakshman-TE-SDN}, need to be extended, e.g., to become more robust, in order to account for this uncertainty.

\subsection{Factor 2: Need for Autonomous Actions}

Tactical networks will exhibit high level of dynamism and frequent network failures. At the same time, quick reactions will be required.
Unfortunately, the fully centralized control paradigm of SDN, where the controller installs all the forwarding rules, will often result in unacceptably slow policy updates.
Hence, a critical question here is how to make data plane nodes to autonomously and quickly react to network changes, but at the same time preserve the benefits of centralized control.

Table I outlines two approaches for enabling autonomous actions in SMANETs. The first approach is rather straightforward and suggests to use a distributed MANET protocol as-a-backup to SDN control. When link failures are detected, the data plane nodes can run a legacy MANET protocol to re-route traffic through alternative paths. Still, the complex network applications may be managed by the SDN controller, and the MANET protocols can be complementary so as to ensure that basic connectivity is restored.

The second approach is to revisit SDN's philosophy of separating control and data planes, and delegate some level of control to data plane nodes. This can be realized by pre-computing and pushing to data plane nodes locally-executable code fragments that realize control functions. An API to program such functions was proposed in \cite{inspired}. If the data plane nodes are enhanced with the ability to perform statefull forwarding actions, as in OpenState~\cite{openstate}, then these functions can take the form of state-dependent forwarding rules.
%OpenState is an extension of OpenFlow that allows data plane nodes to autonomously apply different forwarding rules depending on their current ``state'' and to make states evolve according to state machines.
An example of this approach is depicted in Figure 5. The SDN controller pro-actively computes and installs state-dependent forwarding rules at node 1 to overcome the failure of the link to node 2. In fact, such statefull data plane actions are promoted by popular SDN programming languages.

The first approach is quite simple to implement. Nevertheless, in practice, the MANET protocol can take long time to converge and be too simple to support complex mission applications. On the other hand, the control delegation approach can achieve instantaneous reaction to link failures, but it requires the computation of the code fragments, which can be a very complex problem (especially in large networks with complex policies). Besides, a common code execution platform on all data plane nodes is needed. Developing a method that is both lightweight and agile in terms of reaction to network changes is still an open research problem.

\section{Discussion and Conclusions}\label{sec:conclusions}

SDN is still at its infancy stage in mobile networks, but there is growing consensus that it can be a game-changing force. Ad hoc networking is probably the most challenging area that SDN can be used, but at the same time one that can benefit the most from this new technology. Exploring this potential is of paramount importance not only for the military tactical networks, but for all services and applications that can be supported by ad hoc wireless networks.

\begin{table}
\begin{center}
\begin{tabular}{ |l|l|l| }
\hline
\bf{Reaction Approach} & \bf{Pros} & \bf{Cons} \\ \hline
\multirow{2}{*}{MANET backup} &  Simplicity
 & Operation constrained by  \\
 & ~ & MANET protocol \\ \hline
\multirow{2}{*}{Delegation of control} & Fast reaction & High complexity \\
 & ~ & Need a common platform \\
\hline
\end{tabular}
\small{\caption{Reaction approaches for SDN data plane.}}
\end{center}
\end{table}

In this paper we focused on tactical MANETs and the additional challenges that coalitional operations raise. We studied a new generation of SDN-enabled MANETs, the SMANETs. It was explained that the centralization that SDN brings can help alleviate the inefficiencies of the decentralized MANET designs of the past. At the same time, SDN can maintain the network agility and autonomicity that tactical operations require. Importantly, SDN offers the means to easily configure and re-configure the nodes based on policy, mission objectives and battlefield condition changes. Hybrid SDN networks are an inescapable intermediate step in the deployment of such systems, and we provided a set of guidelines for their design.

The importance of SDN for military operations is already well understood \cite{dinesh}, but to date the focus has been for larger-scale deployments, e.g., see \cite{sdn-gerla} and references there in. Here we went a step further towards the edge of communication systems. Despite the useful lessons from recent experiments and theoretical studies, there are clearly many open challenges. For example, we need to revisit the discussed mechanisms to address security issues, such as Denial-of-Service (DoS) attacks to SDN controllers and data plane nodes. Besides, extensive testbed-based evaluations are necessary, similar to those presented, in order to identify all possible trade-offs and performance limitations of such systems. 

\section{Acknowledgments}

This research was sponsored by the U.S. ARL and the U.K. MoD under Agreement Number W911NF-16-3-0001. The views and conclusions contained in this document are those of the authors and should not be interpreted as representing the official policies, either expressed or implied, of the U.S. Army Research Laboratory, the U.S. Government, the U.K. Ministry of Defence or the U.K. Government. The U.S. and U.K. Governments are authorized to reproduce and distribute reprints for Government purposes notwithstanding any copy-right notation hereon. K. Poularakis acknowledges the Bodossaki Foundation, Greece, for a postdoctoral fellowship. G. Iosifidis acknowledges support by a research grant from Science Foundation Ireland (SFI) under Grant Number 17/CDA/4760.

\begin{IEEEbiographynophoto}{Konstantinos Poularakis} received his Ph.D. in Electrical Engineering from University of Thessaly, Greece, in 2015. Currently, he is a Post-doc researcher at Yale University. His research interests lie in the broad area of network optimization with emphasis on emerging architectures such as software-defined networks.
\end{IEEEbiographynophoto}

\begin{IEEEbiographynophoto}{George Iosifidis} received his Ph.D. in electrical engineering from University of Thessaly, Greece, in 2012, and is currently the Ussher Assistant Professor in Future Networks with Trinity College Dublin. His research interests lie in the broad area of wireless network optimization and network economics.
\end{IEEEbiographynophoto}

\begin{IEEEbiographynophoto}{Leandros Tassiulas} is the John C. Malone Professor of Electrical Engineering and member of the Institute for Network Science, Yale University. He received the IEEE Koji Kobayashi Computer And Communications Award (2016), the Inaugural INFOCOM 2007 Achievement Award, the INFOCOM 1994 Best Paper Award, the National Science Foundation (NSF) Research Initiation Award (1992), the NSF CAREER Award (1995), the Office of Naval Research Young Investigator Award (1997), and the Bodossaki Foundation Award (1999).
\end{IEEEbiographynophoto}


\begin{thebibliography}{99}

\bibitem{sdn-commag} M. Jarschel, T. Zinner, T. Hossfeld, P. Tran-Gia, W. Kellerer, ``Interfaces, Attributes, and Use Cases - A Compass for SDN'', \emph{IEEE Communications Magazine}, vol. 52, no. 6, June 2014, pp. 210–174.

\bibitem{syrivelis} D. Syrivelis, G. Iosifidis, D. Delimpasis, K. Chounos, T. Korakis, L. Tassiulas, ``Bits \& Coins: Supporting Collaborative Consumption of Mobile Internet'', \emph{in Proc. of IEEE INFOCOM}, 2015, pp. 2146-2154.

\bibitem{dinesh} V. Mishra, D. Verma, and C. Williams, ``Leveraging SDN for Cyber Situational Awareness in Coalition Tactical Networks'', \emph{in Proc. of IST-148 Meeting}, 2016, pp. 1-9.

\bibitem{coalition-survey} T. Gibbons, J.V. Hook, N. Wang, T. Shake, D. Street, V. Ramachandran, ``A Survey of Tactically Suitable Exterior Gateway Protocols'', \emph{in Proc. of IEEE Milcom}, 2013, pp. 487-493.

\bibitem{coalition-virtual} J.N. Wang, A. Narula-Tam, R. Byan, ``Inter-domain Routing for Military Mobile Networks'', \emph{in Proc. of IEEE Milcom}, 2015, pp. 407-412.

%\bibitem{p4} P. Bosshart, D. Daly, G. Gibb, M. Izzard, N. McKeown, J. Rexford, C. Schlesinger, D. Talayco, A. Vahdat, G. Varghese, D. Walker, ``P4: Programming Protocol-Independent Packet Processors'', \emph{ACM SIGCOMM Computer Communication Review}, vol 44, no. 3, July 2014, pp. 87-95.

\bibitem{softcell} X. Jin, L. Erran Li, L. Vanbever, J. Rexford, ``SoftCell: Taking Control of Cellular Core Networks'', \emph{in Proc. of ACM CoNEXT}, 2013, pp. 1-14.

\bibitem{softran} A. Gudipati, D. Perry, L.E. Li, S. Katti, ``SoftRAN: Software Defined Radio Access Network'', \emph{in Proc. of ACM HotSDN}, 2013, pp. 25-30.

\bibitem{openradio} M. Bansal, J. Mehlman, S. Katti, P. Levis, ``Openradio: A Programmable Wireless Dataplane'', \emph{in Proc. of ACM HotSDN}, 2012, pp. 109-114.

\bibitem{cp1} B. Heller, R. Sherwood, N. McKeown, ``The Controller Placement Problem'', \emph{in Proc. of ACM HotSDN}, 2012,, pp. 7-12.

\bibitem{cp2} S. Lange, S. Gebert, T. Zinner, P. Tran-Gia, D. Hock, M. Jarschel, M. Hoffmann, ``Heuristic Approaches to the Controller Placement Problem in Large Scale SDN Networks", \emph{IEEE Transactions on Network and Service Management}, vol. 12, no. 1, 2015, pp. 4-17.

\bibitem{hyperflow} A. Tootoonchian, Y. Ganjali, ``Hyperflow: A Distributed Control Plane for Openflow'', \emph{in Proc. of INM conference}, 2010, pp. 3-3.

\bibitem{pist} K. Poularakis, G. Iosifidis, G. Smaragdakis, L. Tassiulas, ``One Step at a Time: Optimizing SDN Upgrades in ISP Networks'', \emph{in Proc. of IEEE INFOCOM}, 2017, pp. 1-9.

\bibitem{inspired} R. Bifulco, J. Boite, M. Bouet, F. Schneider, ``Improving SDN with InSPired Switches'', \emph{in Proc. of ACM SOSR}, 2015, pp. 1-12.

\bibitem{openstate} G. Bianchi, M. Bonola, A. Capone, C. Cascone, ``OpenState: Programming Platform-independent Stateful OpenFlow Applications Inside the Switch'', \emph{ACM SIGCOMM Computer Communication Review}, vol. 44, no. 2, April 2014, pp. 44–51.

\bibitem{sdn-gerla} J. Nobre, D. Rosario, C. Both, E. Cerqueira, M. Gerla ``Toward Software-defined Battlefield Networking'', \emph{IEEE Communications Magazine}, vol. 54, no. 10, October 2016, pp. 152-157.

\end{thebibliography}
\end{document}